# COMPUTATIONAL SIMULATION AND ANALYSIS OF MAJOR CONTROL PARAMETERS OF TIME-DEPENDENT PV/T COLLECTORS

**Jimeng Shi**

Department of Mechanical and Materials Engineering

Florida International University

Miami, Florida, 33174, USA

**Cheng-Xian Lin**

Department of Mechanical and Materials Engineering

Florida International University

Miami, Florida, 33174, USA

**ABSTRACT**

In order to improve performance of photovoltaic/thermal (or PV/T for simplicity) collectors, this paper firstly validated a previous computational thermal model and then introduced an improved computational thermal model to investigate the effects of the major control parameters on the thermal performance of PV/T collectors, including solar cell temperature, back surface temperature, and outlet water temperature. Besides, a computational electrical model of PV/T system was also introduced to elaborate the relationship of voltage, current and power of a PV module (MSX60 poly-crystalline solar cell) used in an experiment in the literature. Simulation results agree with the experimental data very well. The effects of the time-steps from 1 hour to minute, which is closed to the real time, were also reported. At last, several suggestions to improve the efficiency of PV/T system were illustrated.

## 1. INTRODUCTION

A photovoltaic/thermal hybrid solar system consists of a photovoltaic (PV) system which are used for transforming the illumination of sunshine into electrical energy due to photovoltaic effect, and solar thermal systems (T) which are used for transforming solar energy into heat with a kind of fluid. However, as we know that efficiency of the PV/T system decreases with the increase of temperature of PV module. Hence, in order to improve the overall efficiency of solar conversion, it's evidently significant to control the temperature of PV module reasonably.

Early prior researches on PV/T system were started in the 1970's [1-5]. Afterwards, several researchers began using flat plate collectors for PV/T system [6-7] and conducted analytical studies about the new integrated technology by using air and water to cool the PV/T system [8-10] during the 1980's.

In the next 20 years, researches was conducted by Prakash et al. [11] with flowing air and water below the PV module to reduce system's temperature for increasing the efficiency. They analyzed the transient behavior of a photovoltaic/thermal collector for co-generation of electricity and hot flowing fluid. Besides, a configuration of reducing the temperature of the PV/T module by using flowing air as a cool fluid was still applied by Infield et al. [12]. And a non steady state thermal model by considering the effect of heat capacity of the PV/T system was introduced by Jones et al. [13]. Similar researches were also conducted by Chow et al., Tripanagnostopoulos et al., Zondag et al. [14-16].

In 1999, Huang et al. [17] have made an experiment of an integrated photovoltaic/thermal (PV/T) system with 45L water stored in storage tank. They found the performance of heat collecting plates with tube spacing (W) / tube diameter (D) = 6.2 and 10 isn't very reasonable. So they redesigned the heat collecting plate by using corrugated polycarbonate panel configurations. Based on their experiment, Tiwari et al. [18] built a computational model to evaluate some major parameters on the thermal performance of PV/T collector. But an important heat transfer coefficient, the edge loss coefficient was ignored and the radiative heat transfer coefficient from the cell to the sky was not very accurate in his paper. Besides, the parameters including water temperature and cell temperature were calculated. in every hour. We also conducted a computational simulation with a time difference of an hour [19]. Although such simulations still have a good agreement with experimental data [17], such large time-step results would have lower authority when an almost real time (1 minute) is considered.

Therefore, in this paper we managed to develop a more realistic computational model to predict the thermal and electrical performance of PV/T system. We firstly validated the previous computational model, and then incorporated some improvements above the heat transfer coefficients. Moreover, we calculated every parameter in every minutes, which can be closely seen as real time. In addition, considering the fact that the value of electrical power and thermal energy differs in the form of energy, we also built an electrical model and evaluated the efficiency of PV/T system at different cell temperature and different irradiance.

## 2. NOMENCLATURE

| | |
|---|---|
| $A_c$ | Area of PV module (m$^2$) |
| $a$ | Ideality factor (V) |
| $b$ | Breadth of PV/T module (m) |
| $C_w$ | Special heat capacity of water (J/kg.K) |
| EVA | Ethylene vinyl acetate |
| $F_R$ | Flow rate factor |
| $F'$ | Collector efficiency factor |
| $F''$ | Collector flow factor |
| $G$ | Incident solar intensity (function of time) (W/m$^2$) |
| $h$ | Heat transfer coefficient (W/m$^2$·K) |
| $h_{p1}$ | Penalty factor due to the presence of solar cell (W/m$^2$·K) |
| $h_{p2}$ | Penalty factor due to the presence of interface between Tedlar and working fluid (W/m$^2$·K) |
| $I(t)$ | Incident solar intensity (function of time) (W/m$^2$) |
| $I$ | Current (A) |
| $I_d$ | Diode current (A) |
| $I_{mp}$ | Current at maximum power (A) |
| $I_{ph}$ | A light generated current source (A) |
| $I_{RS}$ | Diode reverse saturation current (A) |
| $I_{sh}$ | Shunt current (A) |
| $I_s$ | Cell saturation of dark current (A) |
| $I_{sc}$ | Short-circuit current (A) |
| $k$ | Boltzmann's constant |
| $K$ | Thermal conductivity (W/m.K) |
| $L$ | Length of PV/T module (m) |
| $M$ | Mass (kg) |
| $\dot{m}$ | Mass flow rate (kg/s) |
| $P$ | Power of PV module (W) |
| $Q_u$ | Rate of useful energy (W) |
| $R_s$ | Series resistance (Ω) |
| $R_{sh}$ | Shunt resistance (Ω) |
| $T$ | Temperature (°C) |
| $U_L$ | Overall heat transfer coefficient from solar cell to ambient through top and back surface (W/m$^2$·K) |
| $U_T$ | Conductive heat transfer coefficient from solar cell to water through Tedlar (W/m$^2$·K) |
| $U_b$ | Overall heat transfer coefficient from bottom to ambient (W/m$^2$·K) |
| $U_e$ | Overall heat transfer coefficient from edge to ambient |
| $U_t$ | Overall heat transfer coefficient from solar cell to ambient through glass cover (W/m$^2$·K) |
| $U_{tT}$ | Overall heat transfer coefficient from glass to Tedlar through solar cell (W/m$^2$·K) |
| $U_{tw}$ | Overall heat transfer coefficient from glass to water through solar cell and Tedlar (W/m$^2$·K) |
| $v$ | Velocity (m/s) |
| $V$ | Voltage (V) |
| $V_{mp}$ | Voltage at maximum power (V) |
| $V_{oc}$ | Open-circuit voltage (V) |
| $W$ | Tube spacing (m) |
| $X$ | Simulated parameter |
| $Y$ | Experimental parameter |

Greek Symbols

| | |
|---|---|
| $\beta_c$ | Packing factor |
| $\alpha$ | Absorptivity (mA/°C) |
| $\sigma$ | Stefan-Boltzman constant |
| $\tau$ | Transmissivity |
| *η* | Efficiency (%) |
| *(ατ)eff* | Product of effective absorptivity and transmittivity |
| △ | Difference |

Subscripts

| | |
|---|---|
| *a* | Ambient |
| *bs* | Back surface |
| *c* | Cell |
| *exp* | Experimental |
| *f* | Fluid flow |
| *f_in* | Fluid in |
| *f_out* | Fluid out |
| *g* | Glass |
| *w* | Water |
| *T* | Tedlar |
| *ref* | Refference |
| *th* | Thermal |
| *i* | Insulation |
| *o* | Overall |

## 3. PV/T EXPERIMENTAL DATA AND SETUP

In order to validate the present computational model, we used the original data extracted from experimental setup by Huang et al. [17]. Responding data showed in Figure 1 was obtained on May 21th, 1999.

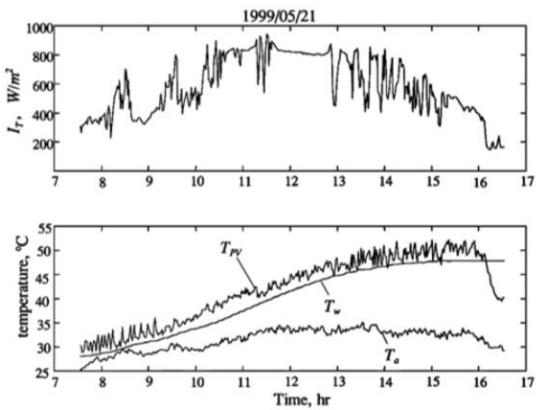

**Figure 1. Experimental results by Huang et al.**

From Figure 1 above we can know specific experimental data including intensity of sunshine, cell temperature in PV model, outlet water temperature in storage tank and ambient temperature in real time, which are compared with the results of computational model in present paper.

Figure 2 shows the schematic diagram of the integrated PV/T system. The setup consists of a commercial PV module (MSX-60 poly-crystalline silicon solar cell) whose total area is 0.516m$^2$ (476mm×1105mm). The poly-crystalline panel is encapsulated in ethylene vinyl acetate (EVA), which is separated from the PV/T collector of the PV/T system by a heat conducting plate made out of copper. There are some thermal grease using between the plate and PV module to get better contact. The system also includes well-insulated pipe from an insulated cylindrical storage tank to the panel. Furthermore, a 3W Direct Current (DC) water pump is installed to force the fluid into the panel and around the whole system. The dimensions of flow channel are 6mm in width, 4mm in height and 0.6mm in thickness. This process of pumping the fluid is called turbulent flow force convection. This is done to further increase the thermal heat gain between the water and the back of the module.

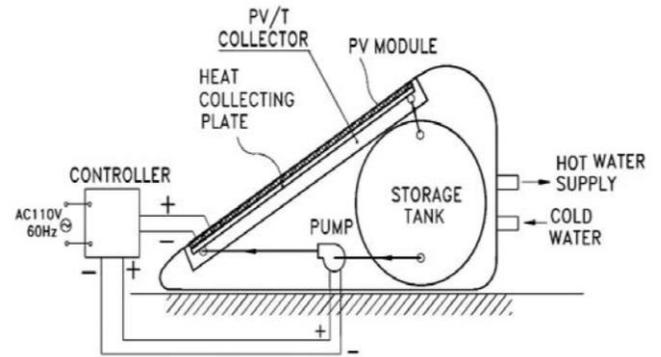

**Figure 2. Schematic diagram of integrated PV/T system**

For seeing the structure of PV/T module clearly, we drew a cross-sectional view of its different components, which is shown in Figure 3. These include glass, ethylene vinyl acetate (EVA), solar cell, Tedlar, water cooling channel and insulation from top to bottom.

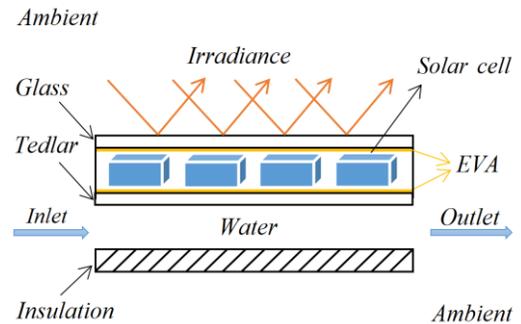

**Figure 3. Cross-sectional view of the PV/T module**

## 4. THERMAL MODEL

In order to guarantee reasonable derivation of following equations of the energy balance in the thermal system, some assumptions are listed below:
- 1D heat conduction has been considered.
- The PV/T system is in a quasi-steady situation.
- The Ohmic losses in the solar cell has been neglected.
- A mean temperature has been supposed across various layers of solar cell.
- The transmissivity of encapsulated material (Ethylene vinyl acetate) is about 100%.
- No stratification in the storage tank which means $T_{fi} \approx T_w$.
- The heat capacities of solar cell material, Tedlar and insulation have been neglected.
- Water flow between the Tedlar and the insulation material is uniform for forced convection.
- The heat capacity of PV/T system is 2918J/K [20] which has been neglected compared with heat capacity of water (188,550J/K) in storage tank.

According to Figure 4, we can clearly analyze the thermal resistance of a PV/T system. It includes conductive resistance between glass, solar cell and Tedlar, convective resistance between Tedlar and fluid flow, radiative resistance between PV/T module and ambient air. This figure can help us clearly understand the heat transfer components that make up the PV/T module with the different colors representing different resistance.

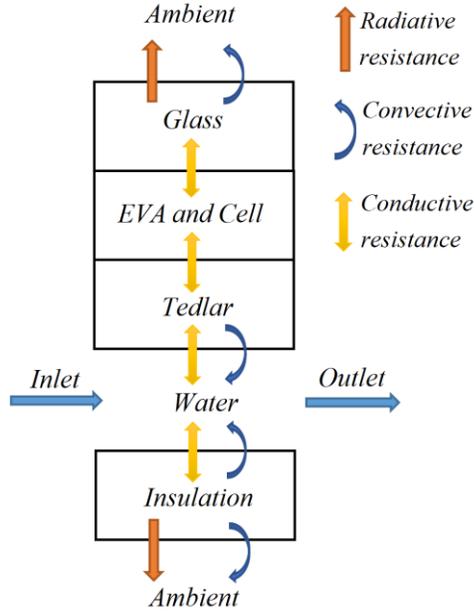

**Figure 4. Thermal resistance of PV/T module**

The heat flow balance in the cooling water channel below the Tedlar along the length of PV/T system is shown in Figure 5, this symbolizes the considerations to calculate the flow balance. Moreover, this figure shows the balance of flow over an elemental length dx = Δx.

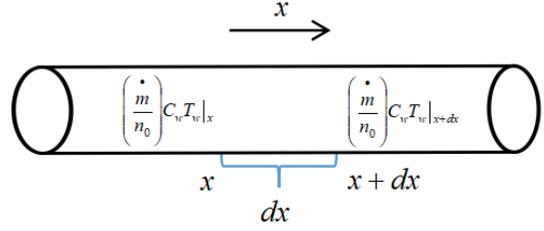

**Figure 5. Heat flow balance in the cooling water channel along the length of PV/T system**

### 4.1 Solar Cells

Equation (1) can be used to calculate the rate of solar energy available on the PV module. This rate of energy available includes the overall heat loss from solar cells to the ambient around module, the overall heat transfer from solar cells to the Tedlar back surface and the rate of electrical energy.

$$\tau_g[\alpha_c I(t)\beta_c + (1-\beta_c)\alpha_T I(t)] \cdot b \cdot dx = [U_t(T_c - T_a) + U_T(T_c - T_{bs})] \cdot b \cdot dx + \eta_c \tau_g I(t)\beta_c \cdot b \cdot dx \quad (1)$$

The amount of thermal energy transferred from solar cells to the Tedlar could be given as follow:

$$U_T(T_c - T_{bs}) = h_{p1}(\alpha\tau)I(t) - U_{tT}(T_{bs} - T_a) \quad (2)$$

$h_{p1}=U_T/(U_t+U_T)$, is the penalty factor caused by the material of solar cells, Tedlar and EVA. It adds a resistance to the thermal model, so it closely represents the actual thermal performance of a PV/T panel.

### 4.2 Tedlar Back Surface

$$U_T(T_c - T_{bs}) \cdot b \cdot dx = h_T(T_{bs} - T_f) \cdot b \cdot dx \quad (3)$$

According to equations (2) and (3), the overall heat transfer from solar cells to the back surface of Tedlar can be equal to the rate of heat transfer from the back surface of Tedlar to the water.

$$h_T(T_{bs} - T_f) = h_{p1}h_{p2}(\alpha\tau)I(t) - U_{tw}(T_f - T_a) \quad (4)$$

$h_{p2}=h_T/(U_{tT}+h_T)$, is the penalty factor caused by presence of the interface between Tedlar and the flowing fluid through the cooling channel. By the way, the detailed calculations for $h_{p1}$ and $h_{p2}$ can be found in the Appendix A with some parameters defined in nomenclature.

### 4.3 Water Flowing Below Tedlar

Due to the available heat transfer below the Tedlar back surface can be taken away by flowing fluid in cooling channel, so the heat transfer should be equal to the mass flow rate of the fluid plus the total heat transfer from the water to the ambient, which is represented by equation (5) below.

$$F'[h_{p1}h_{p2}(\alpha\tau)I(t) - U_L(T_f - T_a) \cdot W \cdot dx] = C_w \cdot \frac{\dot{m}}{n} \frac{dT_f}{dx} \cdot W \cdot dx \quad (5)$$

$U_L = U_b + U_{tw}$, an overall heat transfer coefficient from solar cells to ambient through the back insulation is equal to an overall heat transfer coefficient from water to ambient plus the overall heat transfer coefficient from glass to water through solar cell and Tedlar. And the detailed calculations for $U_b$ and $U_{tw}$ can be found in the Appendix A with some parameters defined in nomenclature.

From the equation (5), we can get the rate of useful energy transfer, $Q_u$, and the instantaneous efficiency at a given time with corresponding climatic data and conditions.

$$\dot{Q}_u = F_R[h_{p1}h_{p2}(\alpha\tau)_{eff}I(t) - U_L(T_{wo} - T_a) \cdot W \cdot dx] \quad (6)$$

$$\eta_i = \frac{\dot{Q}_u}{A_c I(t)} = F_R[h_{p1}h_{p2}(\alpha\tau)_{eff} - U_L \frac{T_{fin} - T_a}{I(t)}] \quad (7)$$

$F_R$, the collector heat removal factor is equal to the collector efficiency factor multiplied by the collector flow factor. The formula is $F_R = F' \times F''$. And the detailed calculations for $F_R$ and $U_{tw}$ can be found in the Appendix A with some parameters defined in nomenclature.

### 4.4 Storage Tank

The available thermal energy from PV/T module to the storage tank is transferred to the water inside the cylindrical tank. The energy balance for the storage tank is given by equation (8) below:

$$\dot{Q}_u = M_w C_w \frac{dT}{dt} + (UA)_T (T_w - T_a) \quad (8)$$

Further, it is important to note that there is no stratification of temperature in the storage tank due to forced mode of operation. Hence one can assume that $T_{fi} = T_w$ as shown in Figure 1. Hence Equations (6) and (8) can be rearranged into a first order differential equation:

$$f(t) = \frac{dT_w}{dt} + mT_w \quad (9)$$

$$m = \frac{(UA)_T + A_c F_R U_L}{M_w C_w} \quad (10)$$

$$f(t) = \frac{A_c F_R h_{p1} h_{p2}(\alpha\tau)_{eff} I(t) + T_a \cdot [(UA)_T + A_c F_R U_L]}{M_w C_w} \quad (11)$$

An analytical solution of Equation (9) can be solved with the assumptions as follows:
- 'm' is considered as constant over the time interval 0–t
- f(t) is considered as an average water temperature over the time interval 0–t

Based on the assumptions above, the analytical solution of Equation (9) with initial condition i.e. $T_w(t=0) = T_{w0}$ is shown in equation (12) below:

$$T_w = \frac{\overline{f(t)}}{m}(1 - e^{-mt}) + T_{w0} e^{-mt} \quad (12)$$

After obtaining water temperature from Equation (12), the temperature of Tedlar back surface of PV module can be obtained from equation (2) and (3):

$$T_{bs} = \frac{h_{p1}(\alpha\tau)I(t) + U_{tT}T_a + h_T T_w}{U_{tT} + h_T} \quad (13)$$

And after gaining the temperature of Tedlar back surface, we can put equation (13) into equation (1) and derive the cell temperature below:

$$T_c = \frac{\tau_g[\alpha_c I(t)\beta_c + (1-\beta_c)\alpha_T I(t)] - \eta_c \tau_g I(t)\beta_c + U_t T_a + U_T T_{bs}}{U_t + U_T} \quad (14)$$

In general, the water temperature and cell temperature with improved model in present paper are computed with equations (12-14), which are shown in Figure 9-12.

### 5. ELECTRICAL MODEL

A PV/T system directly converts the irradiance from solar into electricity via the photovaltaic cell. Hence, we derive the relationship of output current and output voltage (I-V), as well as the relationship of output power and output voltage based on the electrical model. Types of circuits model for PV cell usually are divided below.

### 5.1 Ideal PV Model

An ideal circuit model for PV cell [23] is depicted in Figure 6, which consists of only a single diode connected in parallel with a light generated current source, $I_{ph}$. The output current ($I$) from the PV cell is figured out by applying the *Kirchhoff's Current Law* (*KCL*) can be written as:

$$I = I_{ph} - I_d \tag{15}$$

$$I_d = I_s[\exp(\frac{V}{a} - 1)] \tag{16}$$

$$I = I_{ph} - I_s[\exp(\frac{V}{a}) - 1] \tag{17}$$

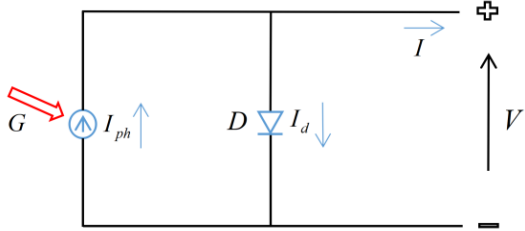

**Figure 6. Ideal circuit model for PV cell**

There are a few things that have not been taken into account in the ideal model and that will affect the performance of a PV cell. Therefore, we add series resistance and shunt resistance in the following models.

**5.2 Practical PV Model with Series Resistance**
In a relatively practical PV cell, there is a series resistance in a current path through the semiconductor material, a diode in parallel and a current source caused by photovoltaic effect. As a series resister ($R_s$), whose effect would become very significant in a PV module that consists of many series-connected cells, and the value of resistance is multiplied by the number of solar cells. The corresponding circuit model is given in Figure 7 as follow.

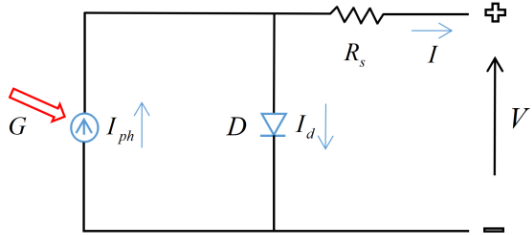

**Figure 7. Circuit model with series resistance for PV cell**

$$I = I_{ph} - I_s[\exp(\frac{V + IR_s}{a}) - 1] \tag{18}$$

**5.3 Practical PV Model with Series and Shunt Resistance**

Although PV model with series resistance is relatively accurate, a more accurate model has been shown in Figure 8, where both of series resistance ($R_s$) and parallel resistances ($R_{sh}$) (also called shunt resistance) [21] are considered. It is a loss associated with a small leakage of current through a resistive path in parallel with the intrinsic device. This can be represented by a parallel resistor ($R_p$) or a shunt resistor ($R_{sh}$). Its effect is much less conspicuous in a PV module compared to the series resistance, and it will only become noticeable when a number of PV modules are connected in parallel for a larger system. Series resistance is usually tiny, which caused by the Ohmic contact between metal and semiconductor internal resistance. However, shunt resistance is usually large and represents the surface quality along the edge, noting that in ideal case $R_s$ is 0 and $R_{sh}$ is ∞ [22].

Therefore, we have chosen the circuit model in Figure 8 for the computational model in the present paper, and we set $R_{sh}$ = 300Ω [23], $R_s$ = 0.5Ω. Then applying *Kirchhoff's Current Law* (*KCL*) again to gain the output current:

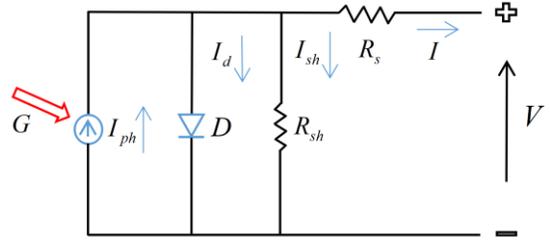

**Figure 8. Circuit model with series resistance and shunt resistance for PV cell**

$$I = I_{ph} - I_d - I_{sh} \tag{19}$$

Diode current is given by equation (16), now we derive the shunt current by equation (20), and then substitute them into equation (19) to get the output current which is shown in equation (21):

$$I_{sh} = \frac{V + IR_s}{R_{sh}} \tag{20}$$

$$I = I_{ph} - I_s[\exp(\frac{V + IR_s}{a}) - 1] - \frac{V + IR_s}{R_{sh}} \tag{21}$$

Based on the typical electrical characteristics of MSX-60 PV cell which are shown in Table 1 [24]. which are measured at $T_c$ = 25°C, G = 1000W/m² by manufacturer, we can determine the series resistance and shunt resistance with the equations (22-27) [25].

Table 1
Typical electrical characteristics of MSX-60

| Typical peak power (Pp) | 60W |
| --- | --- |
| Voltage @ peak power ($V_{pp}$) | 17.1V |
| Current @ peak power ($I_{pp}$) | 3.5A |
| Guaranteed minimum peak power | 58W |
| Short-circuit current ($I_{sc}$) | 3.8A |
| Open-circuit voltage ($V_{oc}$) | 21.1V |
| Temperature coefficient of open-circuit voltage | -(80±10)mV/°C |
| Temperature coefficient of short-circuit current | (0.065±0.015)%°C |
| Approximate effect of temperature on power | -(0.5±0.05)%°C |
| NOCT | 49°C |

Note: NOCT is abbreviation of Nominal Operating Cell Temperature [24].

$$R_s \approx R_{s,ref} = \frac{a_{ref} \ln(1 - \frac{I_{mp,ref}}{I_{sc,ref}}) - V_{mp,ref} + V_{oc,ref}}{I_{mp,ref}} \approx 0.1\,\Omega \quad (22)$$

$$R_{sh} \approx 300\,\Omega \quad (23)$$

According to equation (21), we need to determine the current of photovoltaic cell ($I_{ph}$), cell saturation of dark current ($I_s$), and ideality factor (a). The current of photovoltaic cell mainly depends on the solar insolation and cell's working temperature, which is described as equation (24):

$$I_{ph} = [I_{sc,ref} + K_I(T_c - T_{c,ref})]\frac{G}{G_{ref}} \quad (24)$$

The cell saturation of dark current always keeps changing with cell temperature and $I_{RS,ref}$ is cell reverse saturation current at reference temperature and corresponding solar radiation, which are shown as equation (26) and (27) respectively:

$$I_s = I_{RS,ref} \cdot (\frac{T_c}{T_{ref}})^3 \cdot \exp[\frac{qN_c}{a_{ref}}(1 - \frac{T_{c,ref}}{T_c})] \quad (25)$$

$$I_{RS,ref} = I_{sc,ref} \exp(-\frac{V_{oc,ref}}{a_{ref}}) = 1.565 \times 10^{-6} \quad (26)$$

The ideality factor was introduced by Townsend [26], it is also related to the electrical characteristics of MSX-60 PV cell. The constant a is calculated as below:

$$a = \frac{2V_{mp,ref} - V_{oc,ref}}{\frac{I_{mp,ref}}{I_{sc,ref} - I_{mp,ref}} + \ln(1 - \frac{I_{mp,ref}}{I_{sc,ref}})} = 1.435 \quad (27)$$

Now we have obtained several mathematical equations (19-27), a computational electrical model was inserted into Matlab applications [31] to analyze the relationship between output current, output voltage, solar radiation and cell temperature. In addition, as we know that power should equal voltage multiplied by current. Therefore, the I-V curve and P-V curve with different cell temperatures and irradiance and maximum power point could be obtained [27].

For comparing computational results in the present paper with the experimental data, a root mean square percentage deviation (RMS) can be evaluated by equation (28):

$$RMS = \sqrt{\frac{\Sigma(100 \times \frac{X_{sim,i} - Y_{exp,i}}{X_{sim,i}})^2}{n}} \quad (28)$$

## 6. THE TOTAL EFFICIENCY OF PV/T MODULE

A PV/T system is a combination of photovoltaic (PV) and solar thermal (T) systems which produces both electricity and heat from one integrated system. Therefore, we calculate the thermal and electrical efficiency separately.

**6.1 Thermal efficiency**

With the continuous solar radiation, lots of heat will be generated from the PV/T system, we can calculate the thermal efficiency of improved model using the following formula:

$$\eta_{th} = \frac{M_w C_w (T_f - T_{fin})}{60 \cdot A_c \cdot \Sigma I(t)} \quad (29)$$

We can get the initial water temperature as 28°C, and $\Sigma I(t)$ is the sum of intensity every minute from 8h in the morning to 15h in the afternoon. Some hourly data is shown in Table 2. $T_f$ is the outlet water temperature, 45.3905°C. Other parameters can be found in Appendix A. At last, the overall thermal efficiency is calculated as 40.78%.

Table 2
Hourly temperature data with improved computational model

| Time (h) | $T_{water}$ (°C) | $T_{back\ surface}$ (°C) | $T_{cell}$ (°C) |
|---|---|---|---|
| 8 | 28.7591 | 29.1523 | 31.6681 |
| 9 | 30.3241 | 30.7025 | 33.1204 |
| 10 | 32.3942 | 32.9759 | 36.6840 |
| 11 | 35.3451 | 36.2220 | 41.8202 |
| 12 | 38.3779 | 39.2374 | 44.7147 |
| 13 | 40.7730 | 41.3957 | 45.2925 |
| 14 | 43.4692 | 44.2201 | 48.8955 |
| 15 | 45.3905 | 45.8724 | 48.7740 |

**6.2 Electrical efficiency**

In order to find out the energy conversion efficiency of the solar panel, the following parameters were measured, such as the output power in terms of output current and output voltage, the panel surface temperature and real-time solar radiation intensity (W/m$^2$). In addition to ambient temperature, the inlet and outlet temperature of water flow and water flow rate were measured. The photoelectric conversion efficiency is calculated as:

$$\eta_e = \frac{V_{mp} \cdot I_{mp}}{A_c \cdot I(t)} \times 100\% \tag{30}$$

Therefore, according to the electrical characteristics in Table 1, the energy conversion efficiency of the solar panel at the maximum power should be 15.54%. Although the energy conversion efficiency was given a constant, 9%, in Huang at el. [17], as we known that it should change with the change of output current, output voltage and intensity at any time.

**6.3 The overall efficiency of PV/T system**

For evaluating the PV/T systems, the total efficiency $\eta_o$ is defined in Equation (31), which is equal to the sum of thermal efficiency and electrical efficiency:

$$\eta_o = \eta_{th} + \eta_e \tag{31}$$

From the above equation we can easily find the overall energy efficiency of a PV/T system is always greater than the single thermal efficiency of a solar collector or of the electrical efficiency of a PV module.

**7. MODEL IMPROVEMENTS**

Since the intensity of sunlight fluctuates every minute, if just extracting the hourly data at integral point, what we get from the thermal model is not close to the original experimental situation. So the first improvement in this paper is to reduce the time difference to calculate water and cell temperature, we extract data in every minute with a software called as GetData Graph Digitizer [32], which is so closed to the real time.

Although for most collectors, the evaluation of edge losses is complicated, another improvement is that the edge loss coefficient represented by equation (32) has been considered and included to the overall loss coefficient $U_L$ during the computation of the thermal model, so that we can get more accurate collector overall loss coefficient $U_L$.

$$U_e = \frac{(UA)_{edge}}{A_c} \tag{32}$$

The last improvement is the correction of radiative heat transfer coefficient from the cell to the sky [28], which is related to the overall heat transfer coefficient $U_t$. It has been included as equation (33) and (34) below.

$$T_{sky} = 0.0375636\ T_a^{1.5} + 0.32 T_o \tag{33}$$

$$h_{rad} = \sigma\varepsilon(T_c^2 + T_{sky}^2)(T_c + T_{sky})\frac{T_c + T_{sky}}{T_c + T_{sky}} \tag{34}$$

**8. RESULTS & DISCUSSION**

**8.1 Thermal model**

The experimental results obtained by Huang et al. [17] were used to validate the results produced by the improved computational model in the present paper. The major design parameters for the validation process consist of the different loss coefficients and the specification within the PV/T panel are represented in Appendix A.

Several overall loss coefficients in the improved computational model are shown in equations (35-37):

$$U_L = U_{tw} + U_b + U_e \tag{35}$$

$$U_{tw} = \frac{h_T U_{tT}}{h_T + U_{tT}} \tag{36}$$

$$U_b = [\frac{L_i}{K_i} + \frac{1}{h_i}]^{-1} \quad (37)$$

Where $U_L$ is the overall heat transfer coefficient of the back, edge and top loss coefficients. The back loss coefficient given in equation (37) consists of the thermal resistance of the insulation and the convective heat transfer coefficient on the PV module's back. Additionally, the edge loss coefficient given in equation (32) is based on the collector area. Equation (36) is vital because it indicates the heat transfer which is produced between water in the channel and top layer of PV module.

With the help of equation (12), we can calculate water temperature $T_w$, and put it into equation (13) to get temperature of Tedlar back surface $T_{bs}$, and then we can derive cell temperature from equation (14) which are shown in Figure 9 and Figure 10.

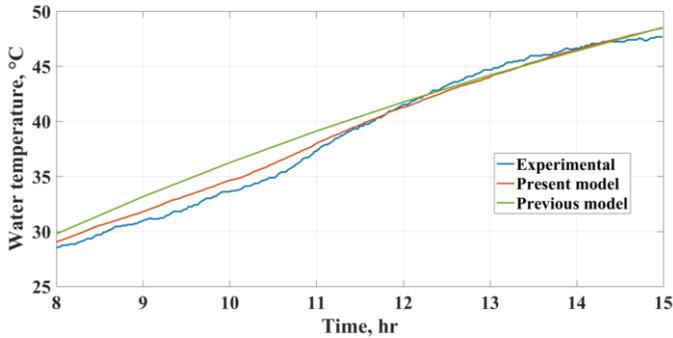

Figure 9. Computational water temperature compared to experimental data

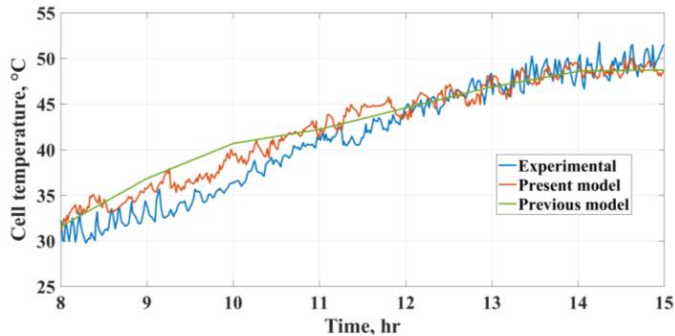

Figure 10. Computational cell temperature compared to experimental data

From figure 9 and figure 10, we can easily find both computational model are valid. However, the time step size is 1 minute in the present model while it's 1 hour in the previous model. Therefore, corresponding results in the present paper are more according with the original experiment than the previous one [18][19]. That's why we obtain a more accurate collector overall loss coefficient $U_L$ by considering the edge loss coefficient $U_e$. With the improved computational model in present paper, a root mean square percentage deviation (RMS) of Figure 9 and Figure 10 are 2.01% and 5.09% while they are 5.87% and 7.22% in precious model respectively.

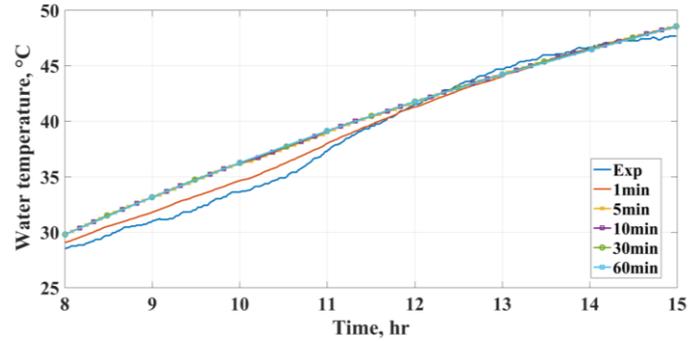

Figure 11. Computational water temperature in difference time step size

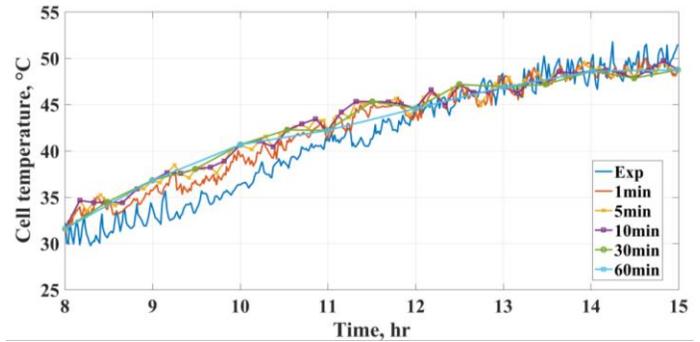

Figure 12. Computational cell temperature in difference time step size

For figuring out the effect of the time step size for calculation to water and cell temperature of PV/T system, we obtain the Figure 11 and Figure 12 in different time step size.

This temperature over the cell is being compared to the theoretical results from the improved model and the original experimental data in different time step size. There is a reasonable agreement between theoretical results by the new and improved model and the experimental results previously published in the literature. That's why we calculate the water temperature and cell temperature in a small time step size, which is almost the real time. A little pity is that even if time difference has been set as 1 minute, the theoretical temperature trend isn't totally satisfying. However, the leap progress in this paper is the cell temperature appears a fluctuating trend like the experimental curve, especially time points appearing maximum and minimum values are similar to those in Huang at el. experiment.

Figure 13 shows the thermal efficiency of PV/T system in real time. We can find it continuously fluctuate with difference time during a day.

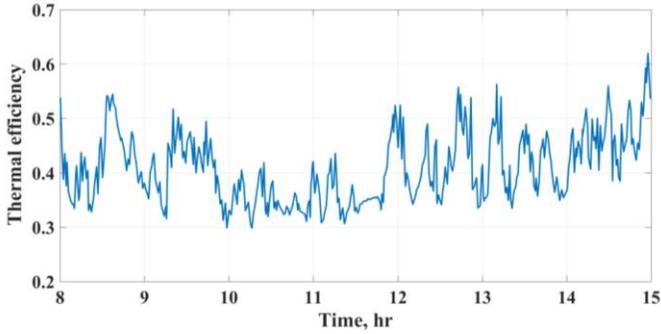

**Figure 13. Thermal efficiency in real time**

**8.2 Electrical model**

The performance of typical MSX-60 modules is described by the I-V curves and electrical characteristics Table 1. Figure 14 is the I-V curve of MSX-60 solar cell provided by manufacturer [24]. In present paper we compared the results in Figure 15 with original data in Figure 14 and extended to analyze the relationship between power and voltage which is shown in Figure 16.

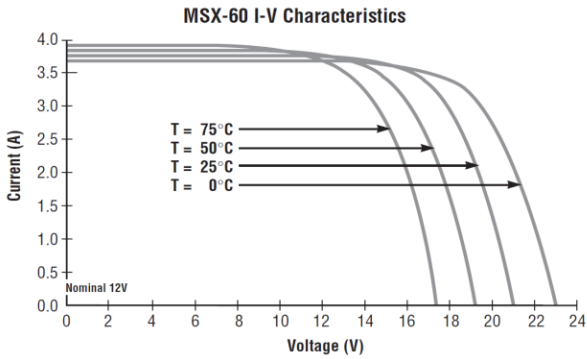

**Figure 14. MSX-60 I-V characteristics**

We use equations (21-27) embodying the computational electrical model, we derive the I-V and P-V curves of MSX-60 solar cell plotted at different temperature levels and with 1000 W/m² irradiance, which is shown as Figure 15, 16 respectively.

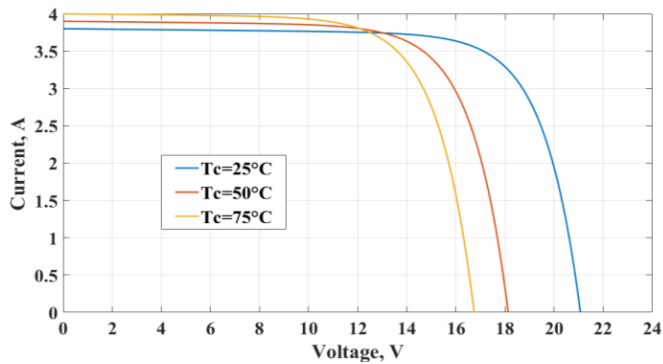

**Figure 15. Computational I-V curve of MSX-60 solar cell plotted at different temperature levels and 1000W/m²**

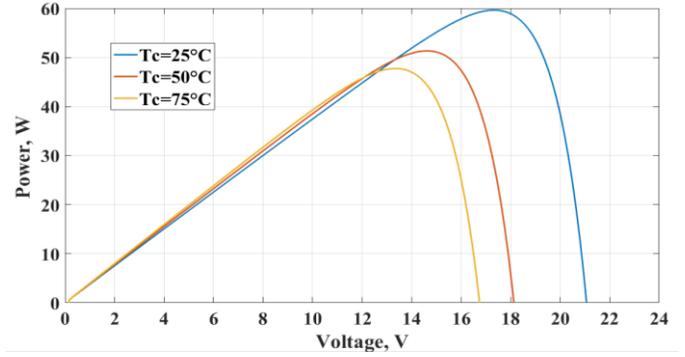

**Figure 16. Computational P-V curve of MSX-60 solar cell plotted at different temperature levels and 1000W/m²**

Figure 15 shows computational I-V characteristics have a good agreement with those provided by manufacturer. Figure 15 and 16 indicate when irradiance is given, the output current slightly increases and operating voltage decreases with increase of cell temperature. In addition, the current and voltage at maximum power decreases with increase of cell temperature.

We use equations (21-27) embodying the computational electrical model, we get the I-V and P-V curves of MSX-60 solar cell plotted at different illumination levels and 25 °C, which is shown in Figure 17 and Figure 18.

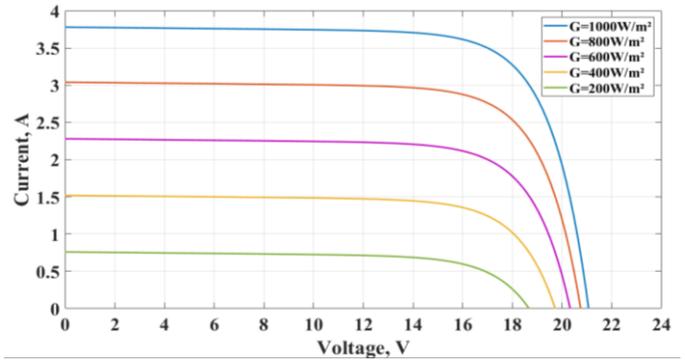

**Figure 17. Computational I-V curve of MSX-60 solar cell plotted at different irradiance levels and 25 °C**

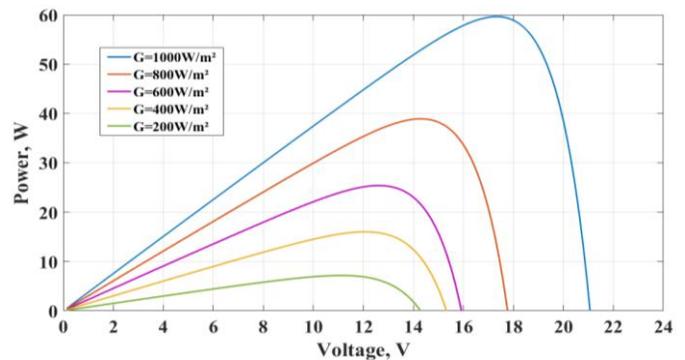

**Figure 18. Computational P-V curve of MSX-60 solar cell plotted at different irradiance levels and 25 °C**

From two figures above, we can easily find the current and power output of PV module are approximately proportional to illumination intensity. The current and voltage at the maximum power increases with the increase of cell temperature.

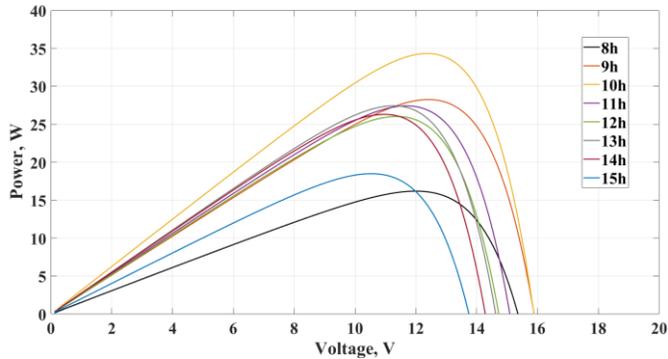

**Figure 19. P-V curve at the corresponding irradiance and cell temperature in hour**

Based on the hourly cell temperature and irradiance, we firstly plotted the P-V curve at the corresponding time. From Figure 19 we can easily get the output current and voltage at maximum power, we obtain the hourly electrical efficiency by using equation (29) which is shown in Figure 20. We can find the electrical efficiency decreases after 10h in the morning with the increase of cell temperature increases.

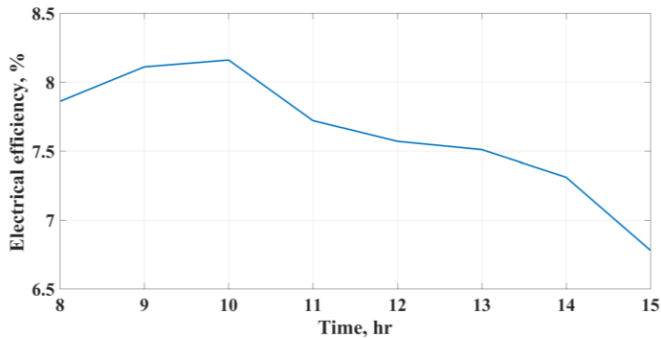

**Figure 20. Electrical efficiency at the corresponding irradiance and cell temperature in hour**

## CONCLUSIONS

A more accurate time-dependent (1 minute as the time difference step size) thermal model was formed to calculate the thermal parameters of a PV/T water collector. Some corrections were added on the heat loss coefficients for improving the computational thermal model. And then we validated the computational results using the improved model with the original experimental data provided by Huang at el. (1999). Further, a computational electrical model was built to analyze the characteristics of MSX-60 solar cell. Finally, some relationship between these parameters were also validated with the typical electrical characteristics provided by manufacturer.

On the basis of present study, the following conclusion has been listed:

1) Because of some correction done to the heat loss coefficients and smaller time step size applied to the present model, it better reflects the change process of water and cell temperature in PV/T system, which are also in better agreement with the experimental data noted in the previous literature. Therefore, we recommend considering more accurate heat loss coefficients listed in nomenclature when designing PV/T systems.

2) Output current and voltage of PV module depend on the irradiation and cell temperature. Moreover, irradiation mainly affects output current, while cell temperature mainly affects the output voltage. Therefore, reasonable utilization of irradiation is really significant to improve electrical efficiency PV/T systems. We should try other way and control time to absorb illumination in order to get more electrical efficiency.

## ACKNOWLEDGMENTS
The authors would like to express gratitude to Beichao Hu, for his useful and important suggestions and help on data analysis.

# APPENDIX A

A.1 Design parameter of PV/T collector and storage tank

| Parameter | Values |
| --- | --- |
| $A_c$ | 0.516 m$^2$ |
| $\triangle \dot{m}$ | 0.016 kg/s |
| b | 0.467 m |
| $C_w$ | 4190 J/kg.K |
| $h_i$ | 5.8 W/m$^2$.K |
| $h_0$ | 5.7+3.8V |
| $h_{p1}$ | 0.8772 |
| $h_{p2}$ | 0.9841 |
| $h_T$ | 500 W/m$^2$.K |
| $K_c$ | 0.039 W/m.K |
| $K_G$ | 1.0 W/m.K |
| Ki | 0.035 W/m.K |
| $K_T$ | 0.033 W/m.K |
| L | 1.105 m |
| $L_c$ | 0.0003 m |
| $L_G$ | 0.003 m |
| $L_i$ | 0.05 m |
| $L_T$ | 0.0005 m |
| $M_w$ | 45 kg |
| $U_b$ | 0.62 W/m$^2$.K |
| $U_t$ | 9.24 W/m$^2$.K |
| $U_T$ | 66 W/m$^2$.K |
| $U_{tT}$ | 8.1028 W/m$^2$.K |
| $(UA)_T$ | 0.44 W/m$^2$.K |
| V | 1m/s |
| $\beta\alpha_c$ | 0.70 ~ 0.85 |
| $\alpha_T$ | 0.50 |
| $\beta_c$ | 0.90 |
| $\tau_c$ | 0.95 |
| $\eta_c$ | 0.09 |

A.2 Some additional formula

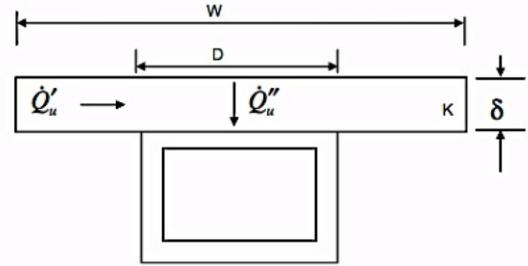

**Figure 11. Schematic view of the panel fined-tube configuration of the flat plate collector**

The finned-tube configuration is illustrated on Figure 14, studied by [30]. Following [29] and using W=0.04 m and D=0.006m, we get:

$$F' = \frac{1}{\frac{WU_L}{\pi Dh} + \frac{WU_L}{\frac{K}{\delta}} + \frac{W}{D+(W-D)F}} \quad (2)$$

$$F'' = \frac{\dot{m}C_f}{A_c U_L F'}[1 - \exp(-\frac{A_c U_L F'}{\dot{m}C_f})] \quad (3)$$

$$F_R = F' \cdot F'' \quad (4)$$

where (1) is the efficiency factor, (2) is the flat plate collector efficiency with $h_T$ being equal to 500 W/m$^2$.°C.

The flow rate factor given by (3) includes the mass flow rate, which was obtained from [30].

$$h_{p1} = \frac{U_T}{U_T + U_t} \quad (5)$$

$$h_{p2} = \frac{h_T}{h_T + U_{tT}} \tag{6}$$

$$U_T = \frac{K_T}{L_T} \tag{7}$$

$$(\alpha\tau)_{eff} = \tau_g [\alpha_c \beta_c + \alpha_T (1-\beta_c) - \eta_c \beta_c] \tag{8}$$